\documentclass[conference]{IEEEtran}
\IEEEoverridecommandlockouts
\usepackage{xcolor}
\usepackage{comment}
\usepackage{diagbox}
\usepackage{booktabs}
\usepackage{colortbl}
\usepackage{multirow}
\usepackage{tabularx}
\usepackage{soul}
\usepackage{wrapfig}
\usepackage[caption=false]{subfig}
\usepackage[font=footnotesize,labelfont=bf]{caption}
\usepackage{dblfloatfix}
\usepackage{enumitem}
\usepackage{tablefootnote}
\usepackage{makecell}
\usepackage{pifont}
\newcommand{\cmark}{\ding{51}}
\newcommand{\xmark}{\ding{55}}
\usepackage{flushend}
\usepackage{booktabs}
\usepackage{cite}
\usepackage{amsmath,amssymb,amsfonts}
\usepackage{algorithm}
\usepackage[noend]{algpseudocode}
\usepackage{graphicx}
\usepackage{textcomp}
\usepackage{url}
\usepackage{bbding}

\usepackage{listings}

\definecolor{lightgreen}{rgb}{0.95, 1.0, 0.99}
\definecolor{darkblue}{rgb}{0.0, 0.0, 0.55}
\definecolor{customlightgray}{rgb}{0.95, 0.95, 0.95} 

\def\BibTeX{{\rm B\kern-.05em{\sc i\kern-.025em b}\kern-.08em
    T\kern-.1667em\lower.7ex\hbox{E}\kern-.125emX}}
\begin{document}

\title{SAGE-HLS: Syntax-Aware AST-Guided LLM for High-Level Synthesis Code Generation\\}

\author{\IEEEauthorblockN{M Zafir Sadik Khan, Nowfel Mashnoor, Mohammad Akyash, Kimia Azar, Hadi Kamali}
\IEEEauthorblockA{\textit{Department of Electrical and Computer Engineering (ECE), University of Central Florida, Orlando, FL 32816, USA} \\
\{mzafirsadik.khan, nowfel.mashnoor, mohammad.akyash, azar, kamali\}@ucf.edu}
}

\maketitle

\begin{abstract}

In today's rapidly evolving field of electronic design automation (EDA), the complexity of hardware designs is increasing, necessitating more sophisticated automation solutions. High-level synthesis (HLS), as a pivotal solution, automates hardware designs from high-level abstractions (e.g., C/C++). However, it faces significant challenges, particularly in design space exploration and optimization. While large language models (LLMs) have shown notable capabilities in code generation, their application to HLS has been limited due to the scarcity of (publicly) available HLS code datasets. Hence, research in this domain has primarily focused on techniques such as prompt engineering and retrieval-augmented generation (RAG). To overcome this limitation, this paper introduces SAGE-HLS, the first-of-its-kind fine-tuned LLM specifically for HLS code generation. Our method includes three key advancements: (i) We implement Verilog-to-C/C++ porting, converting verified and synthesizable Verilog codes into corresponding C, creating a dataset of 16.7K HLS codes; (ii) We implement a fine-tuning strategy, which is based on instruction prompting to code generation guided by abstract syntax tree (AST); (iii) We develop a semi-automated evaluation framework using VerilogEval to assess the functionality of the generated HLS code. Our experiments show that SAGE-HLS, fined-tuned on the QwenCoder (2.5) 7B model, achieves a near 100\% success rate in code synthesizability and a 75\% success rate in functional correctness\footnote{The code and resources related to this work are publicly available at: \url{https://github.com/zfsadik/SAGEHLS}}.

\end{abstract}


\section{Introduction}

High-Level Synthesis (HLS) was introduced to mitigate the inefficiencies of traditional register-transfer level (RTL) design and automate hardware generation from high-level programming languages \cite{coussy2010high, lahti2018we}, aiming to enhance productivity and ease application-specific hardware design space exploration (DSE) \cite{gajski2012high} without hand-coding in hardware description languages (HDL). Over the past decade, HLS has matured, enabling faster iterations and optimizations in hardware design \cite{liu2019accelerating, zhang2021towards, cong2022fpga, cortes2016high}. However, challenges persist in optimizing power, performance, and area (PPA), often requiring extensive manual intervention and expert knowledge \cite{cong2022fpga, shi2023sechls}.

Recent advancements in large language models (LLMs) have shown remarkable capabilities in natural language understanding \cite{brown2020language}, reasoning \cite{jiang2023structgpt}, and particularly code generation (as evidenced by models like GitHub Copilot \cite{dakhel2023github} and CodeGen \cite{nijkamp2022codegen}). In the domain of hardware designs, while LLMs have been applied to tasks such as RTL code generation \cite{thakur2023verigen, blocklove2023chip, liu2023chipnemo, akyash2025rtl++, liu2025craftrtl, akyash2025decortl}, debugging and verification \cite{fang2024assertllm, bhandari2024llm, mashnoor2025llm}, security \cite{akyash2024self, ahmad2024hardware}, etc., their application in HLS has been comparatively limited \cite{akyash2024evolutionary}. A few recent research endeavors have begun to explore the integration of LLMs in HLS \cite{swaroopa2024evaluating, xiong2024hlspilot, liao2024llmsgood, xu2024automated, sheikholeslam2024synthai, xu2024optimizing, collini2024c2hlsc, oztas2024agentic, mashnoor2025timelyhls}, yet these efforts remain narrowly focused on specific aspects of the design process. Some studies have investigated the use of LLMs for optimizing existing HLS code by refining \texttt{pragma} set, or coding styles to enhance synthesizability \cite{xiong2024hlspilot, xu2024optimizing}. Other studies have focused on converting software C/C++ codes into HLS-compatible C/C++ through refactoring or code repair, aiming for code generation \cite{xu2024automated, collini2024c2hlsc}. Some studies also investigated the use of LLM as an alternative for HLS to perform C-to-HDL conversion through direct reasoning \cite{liao2024llmsgood}. 

While these approaches have demonstrated promising results in their scope, they primarily rely on prompt engineering, whether using simple instructions or structured chain-of-thought prompting, all on top of commercially available models, e.g., GPT, Gemini, and Claude \cite{xu2024optimizing, collini2024c2hlsc}. Notably, none of these studies have pursued the development of a domain-specific model by fine-tuning a base LLM for HLS. Additionally, existing LLM-driven HLS approaches face two other key limitations: (i) reliance on raw text-based learning without structural representations, leading to difficulties in handling hierarchy, memory optimizations, and pipelining; and (ii) the lack of a standardized evaluation framework, making results difficult to compare due to variations in benchmarks, synthesis tools, and optimization objectives.

\begin{figure}[b]
\centering
\includegraphics[width=\linewidth]{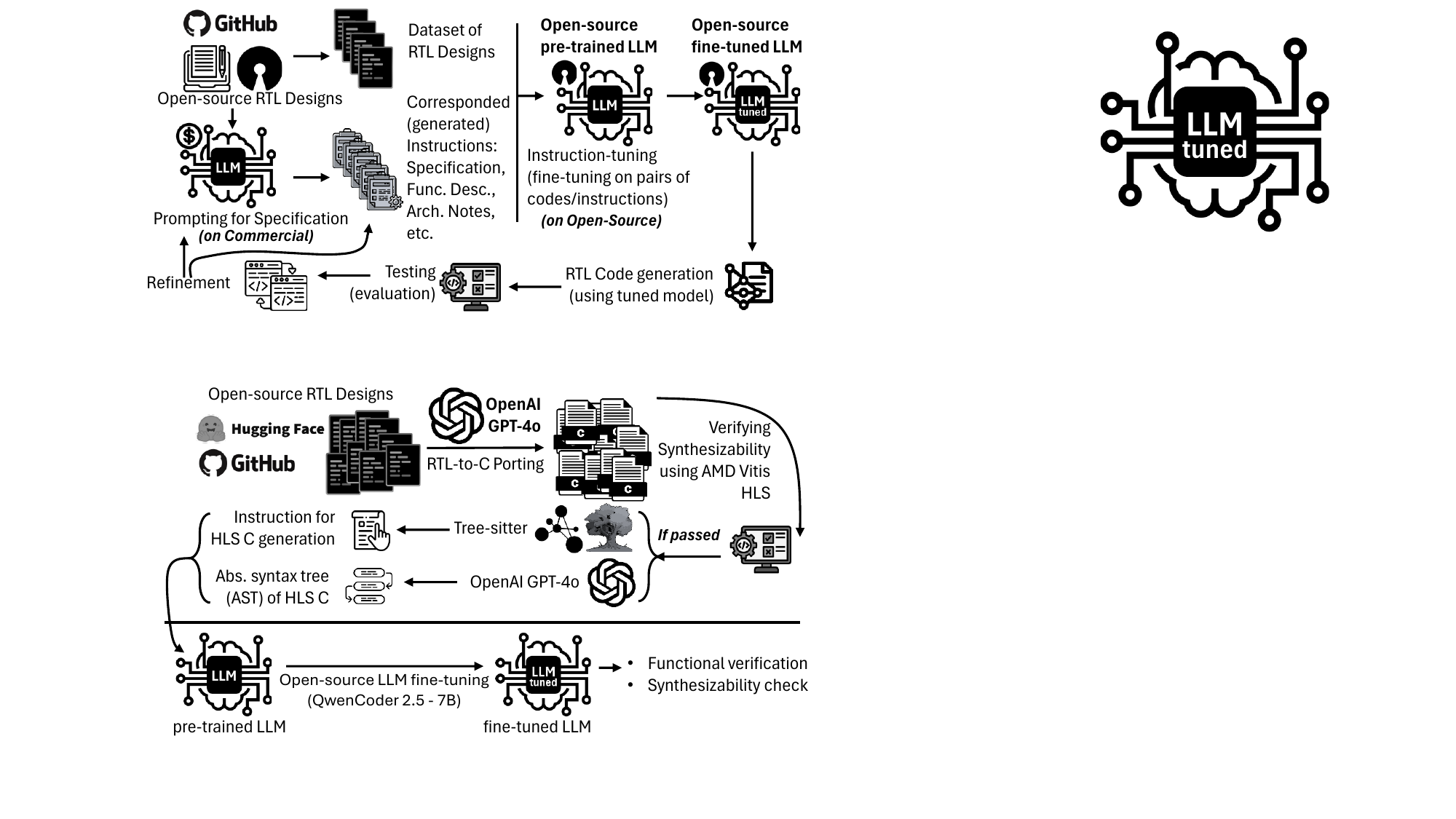}
\caption{Overview of SAGE-HLS: From Code Dateset Preparation using Verilog-to-C Porting, to HLS Code Generation using Fine-Tuned LLM.}
\label{fig:llm_rtl_tuning_overview}
\end{figure}

\begin{table*}[t]
\centering
\scriptsize
\setlength\tabcolsep{1.75pt}
\caption{Overview of Existing LLM-assisted Studies for HLS Code Generation.}
\label{tab:hls_comparison}
\begin{tabular}{@{} l *{6}c @{}}
\toprule
Study & LLM Method & Key Novelty & Evaluation Metrics & HLS Tool Test/Synthesizability \\
\cmidrule(r){1-1}\cmidrule(r){2-2}\cmidrule(r){3-3}\cmidrule(r){4-4}\cmidrule(r){5-5}\cmidrule(r){6-6}

Swaroopa \emph{et al.} \cite{swaroopa2024evaluating} & \makecell{\textit{Prompt engineering} for HLS-C \\ (direct instruction-to-code)} & \makecell{Introducing HLSEval \\ (functional correctness check)} & \texttt{pass@1} functional correctness & Vitis HLS & \\
\cmidrule(r){1-1}\cmidrule(r){2-2}\cmidrule(r){3-3}\cmidrule(r){4-4}\cmidrule(r){5-5}

HLSPilot \cite{xiong2024hlspilot} & Prompt Engineering for design space exploration & Integrating profiling and task pipelining & Area/Latency evaluation & Vitis HLS \\
\cmidrule(r){1-1}\cmidrule(r){2-2}\cmidrule(r){3-3}\cmidrule(r){4-4}\cmidrule(r){5-5}

Liao \emph{et al.} \cite{liao2024llmsgood} & \makecell{\textit{Prompt engineering} for HLS-C, \\ Prompt engineering for C-to-Verilog Synthesis} & LLM as the C-to-Verilog synthesis tool & Area/Latency evaluation & \makecell{Vitis HLS, \\LLM as HLS}\\ 
\cmidrule(r){1-1}\cmidrule(r){2-2}\cmidrule(r){3-3}\cmidrule(r){4-4}\cmidrule(r){5-5}

HLS Repair \cite{xu2024automated} & \textit{Prompt Engineering} for HLS-C Repair & \makecell{LLM-based Step-by-step Repair \\ (using RAG and bit-width optimization)} & Functional pass rate & None \\
\cmidrule(r){1-1}\cmidrule(r){2-2}\cmidrule(r){3-3}\cmidrule(r){4-4}\cmidrule(r){5-5}

SynthAI \cite{sheikholeslam2024synthai} & \textit{Recursive Prompting} for HLS-C & \makecell{Chain of Thoughts (CoT) Prompting \\ using multi-agent LLM} & Generic pass/fail per prompt & None \\
\cmidrule(r){1-1}\cmidrule(r){2-2}\cmidrule(r){3-3}\cmidrule(r){4-4}\cmidrule(r){5-5}

RALAD \cite{xu2024optimizing} & \textit{Prompt Engineering} for \texttt{pargma} management & \makecell{Document Splitting and Retrieval \\ for \texttt{pargma}-based Optimization} & Area/Latency evaluation & Vitis HLS \\
\cmidrule(r){1-1}\cmidrule(r){2-2}\cmidrule(r){3-3}\cmidrule(r){4-4}\cmidrule(r){5-5}

C2HLSC \cite{collini2024c2hlsc} & \textit{Prompt Engineering} for HLS C & \makecell{Automated software C to HLS-C \\ (using iterative LLM-based refactoring)} & \makecell{Functional correctness, \\ Area/Latency evaluation} & Catapult HLS\\
\cmidrule(r){1-1}\cmidrule(r){2-2}\cmidrule(r){3-3}\cmidrule(r){4-4}\cmidrule(r){5-5}

Agentic-HLS \cite{oztas2024agentic} & \textit{Prompt Engineering} for HLS Code Evaluation & \makecell{Agentic reasoning with \\ hierarchical graph embeddings (GNN)} & Area/Latency Evaluation & None \\ 
\cmidrule(r){1-1}\cmidrule(r){2-2}\cmidrule(r){3-3}\cmidrule(r){4-4}\cmidrule(r){5-5}

\makecell{\textbf{SAGE-HLS}\\ \textbf{(Proposed)}}  & \textbf{Fine-tuning code-based LLM for HLS-C} & \makecell{\textbf{Verilog-to-C for DB generation}\\ \textbf{AST-based context-aware fine-tuning}} & \makecell{\textbf{Functional correctness}\\ \textbf{\texttt{Pass@1/5/10}}}  & \textbf{Vitis HLS} \\

\bottomrule
\end{tabular}
\end{table*}


To address these limitations, in this paper, we introduce SAGE-HLS, a novel approach that moves beyond prompt engineering by fine-tuning a base model for HLS code generation to improve the quality (functional correctness) and synthesizability of LLM-generated HLS code (see Fig. \ref{fig:llm_rtl_tuning_overview}). By integrating instruction-based learning and abstract syntax tree (AST) representations, SAGE-HLS is built on the basis of structured learning, leveraging both semantic and syntactical perspectives of coding to enhance automation, robustness, and correctness. Our main contributions are as follows: 

\begin{enumerate}[leftmargin=*]
    \item We construct a large-scale dataset of 20,000 HLS-C codes by reversely converting verifed and synthesizable Verilog to C/C++. This Verilog-to-C porting, conducted by GPT-4, creates a semi-synthetic data useful for fine-tuning. 
    \item We fine-tune a pre-trained model (QwenCoder 7B model \cite{yang2024qwen2}) using instruction-based learning, initially mapping instructions to code, then subsequently enhancing it with AST representations of HLS codes, to capture functional dependencies and improve synthesis outcomes. 
    \item We engage a semi-automated evaluation framework (by using AMD Vitis HLS and VerilogEval \cite{Liu2023verilogeval}) to measure both synthesizability and functional correctness, reflecting the efficiency of the SAGE-HLS model. 
\end{enumerate}

\section{Background and Related Work}

\subsection{LLMs for RTL Code Generation}

Recent advancements in transformer-based hardware design automation have demonstrated the potential of LLMs in various EDA tasks, such as scripting \cite{liu2023chipnemo}, error diagnosis \cite{chang2024data}, and AI-driven design assistants \cite{wu2024chateda}. Among them, numerous studies have focused on fine-tuning and pre-training LLMs for RTL code generation. Early efforts, such as VeriGen \cite{thakur2023verigen}, compiled datasets from GitHub and textbooks but suffered from inconsistencies due to insufficient pre-processing, resulting in frequent syntax errors in generated Verilog code. To improve data quality, RTLCoder \cite{liu2024rtlcodera} introduced RTL-specific keyword extraction to synthesize code-instruction pairs, yet its reliance on GPT-3.5 limited diversity in generated code. Addressing this, OriGen \cite{cui2024origen} employed code-to-code augmentation and self-reflection mechanisms, enabling dataset expansion with syntactically varied but semantically equivalent Verilog while refining code through compiler feedback loops. 

Building upon these advancements, BetterV \cite{pei2024betterv} refined Verilog generation by aligning Verilog semantics with C-like program structures to enhance LLM comprehension, introducing discriminative generation techniques for PPA optimization. AutoVCoder \cite{gao2024autovcoder} tackled domain-specific accuracy and diversity using a two-stage fine-tuning process and retrieval-augmented generation (RAG) for improved correctness. 

Moving beyond generation, CodeV \cite{zhao2024codev} shifted focus to Verilog summarization, curating a dataset of 165K modules to generate high-quality code-description pairs for training. CraftRTL \cite{liu2025craftrtl} integrated state-transition diagrams, Karnaugh maps, and waveforms to enhance structured reasoning in LLMs, while MAGE \cite{zhao2024mage} applied multi-agent reinforcement learning for RTL optimization.

\begin{figure*}[t]
\centering
\includegraphics[width=\linewidth]{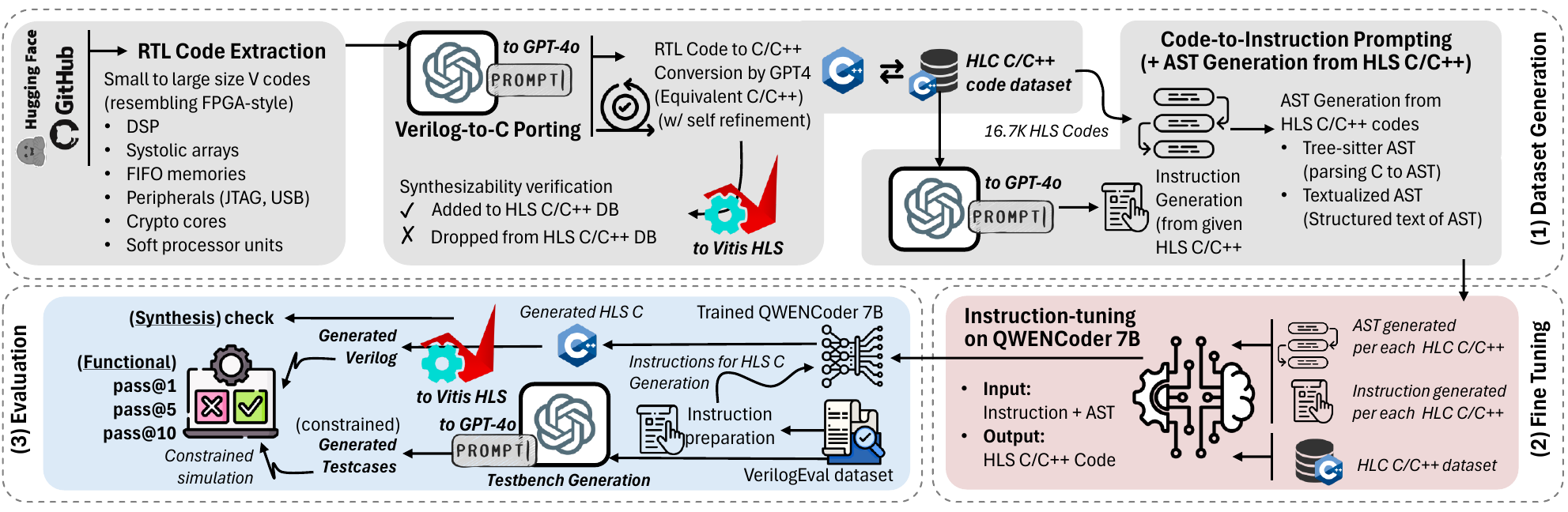}
\caption{Overview of SAGE-HLS: Fine-tuning an LLM for HLS-C generation leveraging AST.}
\vspace{-15pt}
\label{fig:llm_rtl_tuning}
\end{figure*}

\subsection{LLMs for HLS (C/C++) Code Generation}

Unlike RTL designs, which explicitly define cycle-accurate concurrent behaviors, HLS-C introduces hierarchical (sequential) transformations, compiler-driven optimizations, and synthesis constraints that are not explicitly encoded in the text representation of the HLS code. Liao et al. \cite{liao2024llmsgood} provide a broad evaluation of LLM-based approaches for HLS, benchmarking their effectiveness against standard HLS tools like Vitis HLS. Their study highlights that while LLMs can translate C code into hardware descriptions, their performance in terms of power, area, and timing efficiency remains suboptimal due to their lack of explicit structural guidance. 

Similarly, HLSPilot \cite{xiong2024hlspilot} proposes an LLM-driven methodology that integrates profiling, kernel extraction, and DSE to optimize C-to-HLS conversion, demonstrating that LLMs can generate competitive HLS designs when guided with appropriate synthesis constraints.

A key challenge in HLS-based LLM generation is automating the transformation of generic C/C++ programs into synthesizable HLS-C, as explored by C2HLSC  \cite{collini2024c2hlsc}, which investigates whether LLMs can refactor software-like C programs into hardware-compatible representations. Their approach suggests that while LLMs can perform basic transformations, they struggle with hierarchical designs, memory optimizations, and function refactoring, requiring additional preprocessing steps. Xu et al.  \cite{xu2024automated} expand on this challenge by introducing an automated program repair framework that uses retrieval-augmented generation (RAG) to correct C/C++ programs for HLS synthesis, addressing common pitfalls such as dynamic memory allocation, recursion, and improper data types.

Beyond syntax-level generation, recent works have explored multi-agent structured reasoning for HLS. SynthAI employs a structured decision graph with ReAct agents and Chain-of-Thought (CoT) prompting to decompose complex HLS design tasks into manageable subproblems, improving design modularity and synthesis efficiency \cite{sheikholeslam2024synthai}. Agentic-HLS further extends this concept by integrating agentic reasoning into the HLS flow, using graph-based representations to optimize performance predictions and pragma selection \cite{oztas2024agentic}.

Despite advancements, prior studies rely on raw text-based learning, limiting their ability to capture structural dependencies in HLS designs. Our work bridges this gap by integrating AST representations, enabling LLMs to better understand function dependencies, loops, and memory access patterns for more synthesis-friendly HLS code generation. 

\subsection{LLMs for Structural Data Analysis}

With the rise of LLMs, researchers have explored ways to incorporate graph-structured data into LLM inputs, either as embeddings for in-context learning or as structured prompts to improve reasoning over complex relationships \cite{sun2023think}. Given that hardware design inherently involves structural dependencies, such as control flow graph (CFG), data flow graph (DFG), and hierarchical relationships, effective integration of these structural elements into LLM reasoning could be crucial for improving synthesis and optimization tasks.

Several studies have investigated encoding graph information into LLMs. Fatemi et al. \cite{fatemi2023talklikegraph}
 examined the impact of different textual encoding schemes for graph representations, showing that LLM performance in graph reasoning tasks is highly sensitive to encoding strategies, task complexity, and graph structure. Perozzi et al. \cite{perozzi2024letgraph}
 introduced GraphToken, a parameter-efficient embedding method that enhances LLM reasoning by learning structured graph representations instead of relying on textual encoding. Their work suggests that augmenting prompts with explicit graph embeddings significantly improves task performance. Alternatively, GraphLLM \cite{chai2023graphllm} integrates graph learning models directly with LLMs, employing a graph transformer to process graph structures more efficiently, thereby improving both accuracy and scalability.

\section{Proposed Model: SAGE-HLS}

In SAGE-HLS, we aim to address the challenges in LLM-based HLS code generation through three key stages:

\begin{enumerate}[leftmargin=*]
    \item \textbf{\textit{Synthesis-friendly Dataset Creation}}: With the scarcity of reliable (likely synthesizable) HLS code across varying sizes and complexities, from small building blocks (e.g., filters, memory management units (MMUs), crypto cores, etc.) to large-scale designs (e.g., AI accelerators), we have created a dataset by posting verified and synthesizable Verilog code to C/C++, ensuring a high-quality, synthesis-friendly collection of HLS code sutiable for fine-tuning. 
    \item \textbf{\textit{AST Generation from HLS Codes}}: We extracted structural information from the HLS-C code to construct ASTs that encode hierarchical and functional dependencies.
    \item \textbf{\textit{Model Fine-Tuning}}: Using both textual and AST-based representations, we will fine-tune an LLM to enhance their capability in generating synthesis-aware HLS-C code. This dual-representation strategy enables the model to grasp both the semantic and structural properties of the code.
\end{enumerate}

These stages, explained in this section,  collectively enable the model to capture both semantic and structural properties for more robust and efficient HLS-C code generation.

\subsection{Verilog-to-C Porting: Creating and Filtering HLS-C}

To construct a diverse and high-quality dataset for HLS code generation, we first gathered 19K Verilog designs from open-source repositories, e.g., GitHub and Hugging Face, covering a wide range of circuit architectures, including arithmetics and cryptographic cores\footnote{Given that HLS predominantly targets FPGA-based designs, our focus is on array-style codes, e.g., systolic arrays, DSP engines, etc..}. Although there is an abundance of C/C++ codes on open repositories, e.g., GitHub, most of it is software-oriented and thus not directly suitable for HLS. Only a small portion of general-purpose C/C++ codes is written in a synthesizable, hardware-oriented manner style compatible with HLS tools. Using the above RTL implementations as a reference, we employ GPT-4o to generate corresponding HLS-C code and natural language instructions (see Fig. \ref{fig:llm_rtl_tuning} and Listing \ref{list:c2verilog_prompt}), ensuring that the generated HLS-C adheres to high-level programming paradigms while maintaining functional equivalence with the original Verilog designs. 

\begin{lstlisting}[caption={C-to-Verilog Porting using GPT-4 Prompt Engineering.}, label={list:c2verilog_prompt}]
SYSTEM_MSG = '''
You are an expert in hardware design, Verilog, and High-Level Synthesis (HLS). The task is to assist users in converting Verilog into functionally equivalent HLS-C code that can be synthesized using tools like Vivado HLS or Catapult HLS.

When given a Verilog code, you must always generate two outputs:

(1) Equivalent HLS Code: Convert the Verilog into HLS-C while maintaining equivalent functionality. Ensure that the generated HLS-C defines the top-level module as top_module, serving as the entry function. Apply necessary #pragma HLS directives for optimizations, e.g., loop unrolling, pipelining, and memory interfaces. Maintain proper data types and bit-widths to preserve accuracy.

(2) Corresponding Prompt: Generate a generic and simple prompt that describes the hardware functionality concisely. The prompt must be structured in a way that any LLM, including smaller models, can generate the correct HLS code without requiring the original Verilog code. Ensure that the generated HLS code always includes a top_module function as the entry point. You must strictly adhere to this format, ensuring clarity and correctness in both outputs. Do not add unnecessary explanations - focus on delivering precise and structured responses.
'''
\end{lstlisting}

Through this step, we ensure that the generated HLS code are functionally equivalent with the original Verilog designs, while it inherently preserves the synthesis-friendly characteristics of the designs. Furthermore, upon manual inspection of the HLS-C code generated by GPT-4o, we observe that pragma directives are added which are valid and provide guidance that is directionally consistent with typical HLS optimization practices. Also, from Listing \ref{list:c2verilog_prompt}, it is evident that no platform or board is specified during the dataset generation. This generates a generic, algorithm-only implementation that can be compiled and simulated in any HLS tool. So, the generated dataset is not limited for a specific board or HLS tool, thus the resulting HLS-C code is synthesizable. To enforce synthesis compatibility, we validated the synthesizability of the generated HLS-C codes using Vitis HLS. This step filtered out non-synthesizable or inefficient implementations, leaving 16.7K verified HLS-C designs that met synthesis requirements.
 
\subsection{AST Generation for AST-based Fine-tuning}

To extract structural information from HLS-C code, we utilize Tree-sitter \cite{tree-sitter}, a widely used incremental parsing framework that efficiently generates ASTs. The AST representation can provide a formal hierarchical decomposition of the program, capturing essential structural properties such as syntactic constructs, function call dependencies, control flow structures, and memory access patterns\footnote{Due to the inherently structured nature of HLS-C, the AST derived from HLS-C provides a more precise and constrained representation of the code's structure compared to traditional software C \cite{ye2022scalehls}.}. 

\begin{algorithm}[t]
\footnotesize
\caption{AST Extraction and Control Flow Analysis}
\label{alg:ast_hls}
\begin{algorithmic}[1]
\Require HLS-C source file $S$
\Ensure Optimized AST $T'_{main}$ and Control Flow Graph $CFG$

\Function{ParseAST}{$S$}
    \State $T \gets \text{Tree-sitter.parse}(S)$
    \State $N_{main} \gets \text{FindNode}(T, \text{"main"})$
    \State \Return $\text{Subtree}(T, N_{main})$
\EndFunction

\Function{OptimizeAST}{$T_{main}$}
    \For{each node $N \in T_{main}$}
        \If{$N \in \text{RedundantNodes}$}
            \State $T_{main} \gets \text{RemoveNode}(T_{main}, N)$
        \ElsIf{$\text{HasSingleChild}(N)$}
            \State $T_{main} \gets \text{CollapseNode}(T_{main}, N)$
        \EndIf
    \EndFor
    \State \Return $T_{main}$
\EndFunction

\Function{AnalyzeControlFlow}{$T'_{main}$}
    \State $CFG \gets \emptyset$
    \For{each node $N \in T'_{main}$}
        \State $handler \gets \text{Handlers}(N)$
        \State $CFG \gets CFG \cup handler$
    \EndFor
    \State \Return $CFG$
\EndFunction

\Function{Handlers}{$N$}
    \State \Return $
    \begin{cases}
        \{(N, T(N)), (N, E(N))\}, & N.type=\texttt{if} \\
        \{(N, L(N)), (L(N), N)\}, & N.type \in \{\texttt{for}, \texttt{while}\} \\
        \{(N, c_i) | c_i \in C(N)\}, & N.type=\texttt{switch} \\
        \{(N, F(N))\}, & N.type=\texttt{function} \\
        \{(N, R(N))\}, & N.type=\texttt{return} \\
        \{(N, Expr(N))\}, & N.type=\texttt{expression} \\
        \{(N, D(N))\}, & N.type=\texttt{declaration} \\
        \{(N, A(N))\}, & N.type=\texttt{assignment} \\
        \{(N, C(N))\}, & N.type=\texttt{call} \\
        \emptyset, & \text{otherwise}
    \end{cases}$
\EndFunction

\Function{Main}{$S$}
    \State $T_{main} \gets \text{ParseAST}(S)$
    \State $T'_{main} \gets \text{OptimizeAST}(T_{main})$
    \State $CFG \gets \text{AnalyzeControlFlow}(T'_{main})$
    \State \Return $(T'_{main}, CFG)$
\EndFunction
\end{algorithmic}
\end{algorithm}

Unlike raw token-based representations, by using ASTs, we leverage deeper structural insights when fine-tuning our LLM. By wrapping our processing pipeline around Tree-sitter, we systematically extract typed tree structures that organize HLS-C code into a context-free grammar representation, allowing for efficient traversal and transformation. This approach preserves key hardware-relevant abstractions, including loop unrolling, function inlining, pipeline directives, and memory partitioning strategies, which are crucial for high-performance hardware synthesis. Additionally, ASTs improves static analysis of code by explicitly capturing data dependencies and control structures, enabling our model to learn how different code components contribute to scheduling, resource allocation, and computational parallelism in an HLS design. Furthermore, AST-based representations provide a layer of abstraction between software-like HLS-C code and it's underlying RTL implementation, helping the model bridge the gap between algorithmic specification and hardware synthesis constraints whether those are pragmas, directives or external tool constraints. Alg. \ref{alg:ast_hls} shows a high-level yet step-by-step implementation of AST, followed by generic optimization, leading to the extraction of the CFG per each HLS-C. 

\subsection{LLM Fine-tuning for HLS-C Generation}

To enable LLM to generate synthesis-friendly HLS-C code, we fine-tune QWENCoder (2.5) 7B \cite{yang2024qwen2}, an advanced open-source model designed for code generation. Our fine-tuning strategy is designed to evaluate the impact of structural information by training two separate variants: (i) a baseline model fine-tuned only on raw HLS-C text (called \textbf{QWEN-HLS}) and (ii) an AST-enhanced model trained with both HLS-C text and its corresponding AST representation (called \textbf{SAGE-HLS)}. In the AST-augmented variant, (from Alg. \ref{alg:ast_hls}) we prepend a serialized AST representation to the input, allowing the model to process hierarchical relationships, function dependencies, and control flow structures that are otherwise lost in token-based representations. This additional structured context provides the model with explicit syntax-awareness which helps it make better synthesis decisions, such as loop transformations, memory optimizations, and pipeline scheduling. Listing \ref{list:training_input} illustrates the input format provided to the base model during the fine-tuning process of the AST-augmented model.

\begin{lstlisting}[caption={HLS Code, Instruction and AST for training SAGE-HLS.}, label={list:training_input}]
HLS_CODE = '''
#include <ap_int.h>
#include <hls_stream.h>
// Top-level function for HLS synthesis
void top_module(
ap_uint<11> v_addr, ap_uint<8>& v_data,
bool v_en, bool& v_rdy
) {// ROM definition and initialization
static const ap_uint<8> rom[2048] = {/*Add ROM data here*/};
// Signal management
#pragma HLS PIPELINE II=1
v_rdy = v_en;
if (v_en) {
v_data = rom[v_addr];
}}
'''

HLS_INSTRUCTION = '''
Create a C++ function named `top_module` that simulates a ROM with the following behavior:
- The top_module accepts 4 inputs: a 11-bit address (`v_addr`), an 8-bit data output reference (`v_data`), a boolean enable signal (`v_en`), and a boolean ready signal output reference (`v_rdy`).
- When `v_en` is high, the ROM outputs the data at the location specified by `v_addr` and sets `v_rdy` to true.
- The ROM should have a size of 2048 entries (addressable using the 11-bit address) with 8-bit data in each entry.
- Use a static array to represent the ROM content and initialize it with some placeholder data.
- Optimize the function by pipelining it with a single initiation interval.
'''

AST = '''
FuncName: top_module, Params: ap_uint<11>, ap_uint<8>, bool, bool
VarTyp: ap_uint<8>
Asgnmnt: v_rdy = v_en
IfStmt: Contn: (v_en)
Then:
Asgnmnt: v_data = rom[v_addr]
'''
\end{lstlisting}

To ensure that the generated HLS-C code is not only structurally correct but also synthesizable with efficient hardware characteristics, we embed pragma annotations during both dataset generation and model fine-tuning. These annotations are selectively added based on AST-guided structural insights such as loop depth, memory access patterns, and function hierarchy. During fine-tuning, the input sequence is a concatenation of (i) instruction prompt, (ii) serialized AST, allowing the model to learn contextual correlations between high-level specifications, structural code properties, and hardware optimization directives. This combination enables the model to better infer where and how to place pragmas. For instance, loop nodes with independent iterations in the AST are often associated with unrolling or pipelining pragmas, while top-level I/O functions receive interface-related annotations. The model learns these associations during fine-tuning through a diverse set of examples where pragma usage varies depending on code topology. This results in HLS-C code that is not only functionally correct but also optimized for performance, as evidenced in our evaluation results.

\subsection{Evaluation on Modified VerilogEval}

\begin{table*}[b]
\footnotesize
\centering
\setlength\tabcolsep{3pt}
\caption{SAGE-HLS Performance vs. Base LLM Model, Showcasing the Impact of Fine-tuning and AST-based Context Enahncent.}
\label{tab:main_comparison_llms}
\begin{tabular}{@{} l *{21}c @{}}
\toprule
\multirow{2}{*}{Evaluated Model} & \multicolumn{3}{c}{Synthesizability Ratio} & \multicolumn{3}{c}{Functional Correctness Ratio} \\
\cmidrule(r){2-4} \cmidrule(r){5-7}
& Synth@1 & Synth@5 & Synth@10 & Pass@1 & Pass@5 & Pass@10 \\
\cmidrule(r){1-1} \cmidrule(r){2-4} \cmidrule(r){5-7}
QWENCoder 7B Pre-trained (\textbf{QWEN Base} Model) & 52.56\% & 61.54\% & 70.51\% & 22.44\% & 38.46\% & 43.59\% \\
\cmidrule(r){1-1} \cmidrule(r){2-4} \cmidrule(r){5-7}
QWENCoder 7B Fine-tuned using \{HLS-C, Instruction\} (\textbf{QWEN-HLS}) & \textbf{\ul{94.87\%}} & 98.72\% & \textbf{\ul{100\%}} & 56.41\% & 67.95\% & 71.79\%  \\
\cmidrule(r){1-1} \cmidrule(r){2-4} \cmidrule(r){5-7}
\textbf{AST-Guide QWENCoder 7B Fine-tuned using \{HLS-C, Instruction\}} (\textbf{SAGE-HLS}) & 92.95\% & \textbf{\ul{100\%}} & \textbf{\ul{100\%}} & \textbf{\ul{57.69\%}} & \textbf{\ul{70.51\%}} & \textbf{\ul{75.64\%}} \\
\bottomrule
\end{tabular}
\vspace{-10pt}
\end{table*}

A key challenge in evaluating HLS-C code is the absence of a standardized benchmark. To address this, we establish a structured evaluation framework derived from VerilogEval \cite{Liu2023verilogeval}, a dataset (followed by simulation-based verification) originally designed for LLM-based RTL verification. However, assuming that HLS-C is synthesized to RTL (e.g., using Vitis HLS), VerilogEval’s direct application to these RTLs is limited by differences in HLS behavior, particularly the timing variations (e.g., introduced by Vitis HLS when targeting FPGA architectures). It prevents one-to-one validation of HLS-generated RTL vs. reference designs. To bridge this gap, we introduce a semi-automated verification methodology that adapts VerilogEval for HLS-C through constrained simulation\footnote{It derives from constrained random verification (CRV) \cite{haedicke2012crave} but generates deterministic stimulus via reference code, existing testbench, and instructions.}. 

Our evaluation begins by transforming VerilogEval’s instructions into an HLS-compatible format. These adapted instructions prompt our fine-tuned LLM to generate HLS-C code. Now via constrained simulation (see Fig. \ref{fig:llm_rtl_tuning}), testbenches from VerilogEval are modified using GPT-4 to incorporate specific constraints that validate functional correctness based on reference code, existing testbench, and instructions. The number of constraints introduced corresponds to the number of test cases defined by VerilogEval for each circuit. The modified testbenches (containing constraints) are then executed using a RTL simulator. If any constraint fails, the system flags the design as incorrect. This structured verification process ensures that our evaluation rigorously assesses the correctness of HLS-generated designs while accommodating the synthesis-driven differences inherent in targeted implementations.

\section{Experimental Setup}

We evaluate our approach by comparing the baseline QWENCoder, text-only fined-tuned model (QWEN-HLS), and the AST-enhanced fine-tuned model (SAGE-HLS). The evaluation phase includes three main steps: (i) \emph{\ul{HLS-C generation}}: We use HLS-aligned VerilogEval instructions as the prompts to our fine-tuned model, which generate corresponding HLS-C code; (ii) \emph{\ul{Synthesizability check}}: We run AMD Vitis HLS on generated HLS-C codes, where the output would be synthesized RTL code generated by Vitis HLS; (iii) \emph{\ul{Functionality correctness}}: We run our semi-automated constraint-based simulation on the HLS-generated RTL, where constraints (pass or fail) show the correctness of the code.    

Our experiments are conducted using QwenCoder (2.5) 7B \cite{yang2024qwen2}, a well-optimized model for code generation. To enable efficient fine-tuning, we apply low-rank adaptation (LoRA), which significantly reduces memory overhead while maintaining fine-tuning effectiveness \cite{hu2022lora}. Also, we utilize 4-bit quantization to minimize the memory footprint and accelerate inference without compromising model performance. The fine-tuning process is configured with a: 

\begin{itemize}[leftmargin=*]
    \item Per-device train batch size of 2;
    \item Gradient accumulation steps of 4; 
    \item One training epoch with a linear learning rate scheduler;
    \item The learning rate of 2e-4;
    \item AdamW 8-bit optimization;
    \item Weight decay of 0.01 (for better generalization);
    \item 5-step warmup, with a seed of 3407 (for reproducibility).
\end{itemize}

To ensure a fair evaluation on the VerilogEval benchmark, we conducted an experiment to verify that our dataset's instructions are not similar to those in VerilogEval. We calculated the ROUGE-L \cite{lin2004rouge} similarity scores between our instructions and those from VerilogEval, and found that all scores were below 0.4. This low similarity confirms that our model learns to generalize rather than memorize specific instruction patterns. For initial dataset collection and creation, including Verilog-to-C porting, instruction generation, and testbench augmentation for constrained simulation, we utilized the OpenAI API\footnote{The total cost for processing the entire dataset amounted to around \$480.}. For the synthesis (using AMD Vitis HLS 2024), we used AMD's Zynq XCZU3EG-SBVA484 MPSoC with standard speed grade and extended temperature range. 

\section{Experimental Results}

Table \ref{tab:main_comparison_llms} shows a comparative analysis of QWENCoder 7B (base, QWEN-HLS, and SAGE-HLS) in terms of synthesizability and functional correctness across multiple 1, 5, and 10 runs (to observing the likelihood of producing a correct result after multiple attempts). First, we check synthesizability, ensuring that the generated code can successfully compile into RTL using Vitis HLS. As shown in Table \ref{tab:main_comparison_llms}, while the pre-trained (base) QWENCoder 7B model achieves low synthesizability scores, both QWEN-HLS and SAGE-HLS demonstrate a significant improvement, achieving nearly perfect synthesizability. Additionally, SAGE-HLS performs similarly to QWEN-HLS in terms of synthesizability\footnote{Only 2 designs (synth@5) and 3 designs (synth@1) showed different synthesizability results between QWEN-HLS and SAGE-HLS.}, suggesting that AST-based structural learning does not significantly improve syntactical formation of the HLS-C for synthesizability.

In terms of functional correctness, the pre-trained QWEN model has the lowest correctness ratios, showing that most of its generated code does not pass functional validation.
While the QWEN-HLS model significantly improves correctness, the AST-Guided SAGE-HLS model consistently outperforms QWEN-HLS in functional correctness, thanks to AST integration, which enables the model to learn structural dependency for more robust (and consistent) code generation. 

\begin{figure*}[t]
\centering
\includegraphics[width=\linewidth]{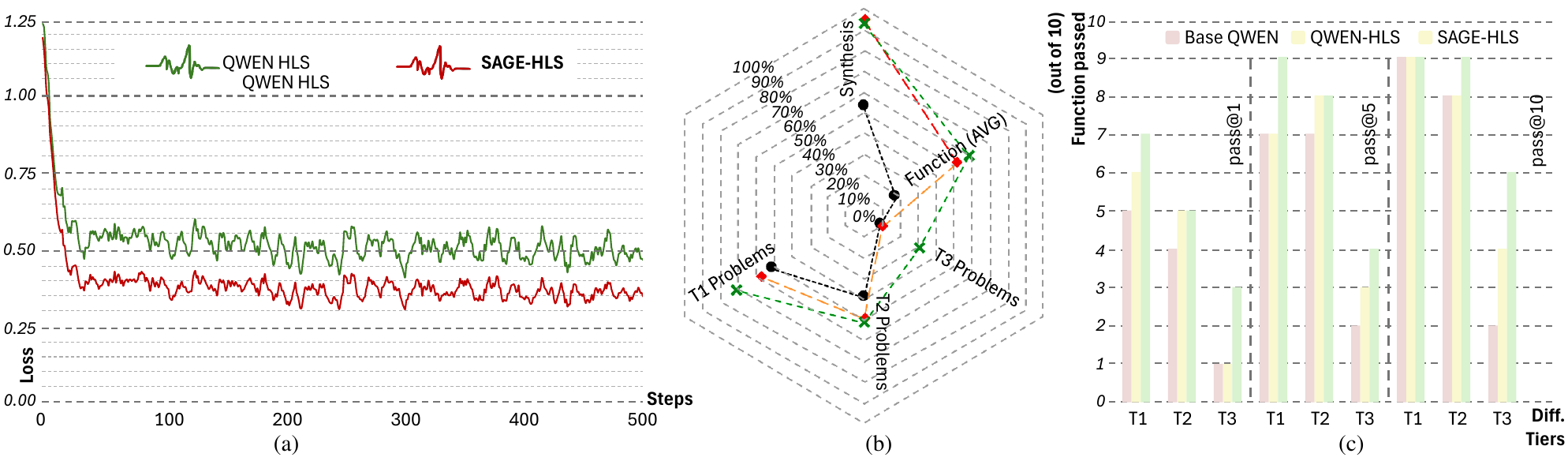}
\caption{Detailed Performance Analysis of SAGE-HLS: (a) Training Loss Comparison between QWEN-HLS and SAGE-HLS; (b) Synthesizability and Functional Correctness on Different Tiers; (c) Detailed Functional Correctness for Pass@1/5/10 on Different Tiers.}
\label{fig:results_loss_distribution}
\end{figure*}

To further evaluate HLS-SAGE, we categorize benchmarks into three difficulty tiers based on the number of characters in the reference Verilog code from VerilogEval: (T1) easy (shorter RTL codes); (T2) intermediate (moderate length of RTL codes); and (T3) hard/complex (large codes, involving loops, deep pipeline, huge state machines, etc.). Tables \ref{tab:synthesis_results} and \ref{tab:pass_results} assess how models handle varying complexity levels in terms of synthesizability and functional correctness. As shown in Table \ref{tab:synthesis_results}, while QWEN base model struggles more with synthesizability (in T2 and T3), even at @5 and @10, synthesizability  is nearly independent of difficulty tiers in trained models\footnote{Almost all circuits from trained models passed synthesizability in 2nd run.}. As shown in Table \ref{tab:pass_results}, all circuits are cases that failed in pass@1. As the number of samples increases to @5 and @10, functional correctness rates improve, where T1 (Easy) benchmarks achieve correctness faster than T2 and T3, and For the most T3 cases, SAGE-HLS consistently performs the best (even for larger benchmarks).

In Fig. \ref{fig:results_loss_distribution}(a), we plot the training loss of two models (i.e., QWEN-HLS and SAGE-HLS) over 500 steps. As shown, SAGE-HLS with the AST-enhanced model (red curve) consistently converges faster, showing a more rapid drop in loss during the initial training phase and maintaining a lower loss value throughout the process compared to the QWEN-HLS without AST (blue curve). Additionally, the SAGE-HLS curve exhibits less fluctuation, suggesting a smoother optimization trajectory. These observations indicate that including AST information provides richer syntactic and semantic context, thereby reducing ambiguity and improving the model’s ability to learn code patterns and structural information more efficiently. Consequently, the model augmented with AST converges to a lower loss, suggesting a more robust understanding of the code and implying that AST-based training can offer significant advantages in specialized language modeling tasks such as HLS-C code generation.

Fig. \ref{fig:results_loss_distribution}(b) summarizes a radar plot comparing the synthesizability and functional correctness of models across different benchmark difficulty tiers (for @1). This plot reflects the average synthesizability and functionality, accompanied with functional correctness of different tiers. As shown, synthesizability (for only one run) is effectively a solved problem for fine-tuned models, achieving near-perfect results. For functional correctness, there exists a significant gap between fine-tuned models and the base one (from $\sim$20\% to $\sim$60\%). This analysis reinforces the necessity of structurally informed models like SAGE-HLS, which consistently outperform standard fine-tuned models (QWEN-HLS) in handling complex functional dependencies in HLS-C generation. 

Fig. \ref{fig:results_loss_distribution}(c) shows a detailed breakdown of functional correctness, showing how many of 10 different designs (per tier) successfully passed functional verification in constrained simulation. The designs are distributed across three difficulty tiers (T1, T2, and T3 --- each 10 separate designs) to illustrate how model performance varies based on circuit complexity. As shown T1 (and almost T2) circuits are easier to generate functionally correct code for, while T3 see a sharp decline. However, with multiple attempts (e.g., @10), T3 shows the biggest improvement, particularly in SAGE-HLS, reinforcing the importance of structural data analysis, such as AST guidance, for more robust HLS-C code understanding.

\begin{table}[b]
\footnotesize
\caption{Synth@1, @5, and @10 (Synthesis) for Selected Benchmarks.}
\centering
\setlength\tabcolsep{4pt}
\begin{tabular}{@{} l *{12}c @{}}
\toprule
\multirow{2}{*}{Benchmark} & 
\multicolumn{3}{c}{QWEN Base} & 
\multicolumn{3}{c}{QWEN-HLS} & 
\multicolumn{3}{c}{\textbf{SAGE-HLS}} \\ 
\cmidrule(r){2-4} \cmidrule(r){5-7} \cmidrule(r){8-10}
& @1 & @5 & @10 & @1 & @5 & @10 & @1 & @5 & @10 \\
\cmidrule(r){1-1} \cmidrule(r){2-4} \cmidrule(r){5-7} \cmidrule(r){8-10}
popcount3  (T1)          & \xmark & \cmark  & \cmark  & \xmark & \cmark  & \cmark  & \xmark & \cmark  & \cmark  \\
\cmidrule(r){1-1} \cmidrule(r){2-4} \cmidrule(r){5-7} \cmidrule(r){8-10}
dff8r (T2)        & \xmark & \xmark  & \cmark  & \xmark & \cmark  & \cmark  & \xmark & \cmark  & \cmark  \\
\cmidrule(r){1-1} \cmidrule(r){2-4} \cmidrule(r){5-7} \cmidrule(r){8-10}
rule110 (T3)      & \xmark & \xmark  & \xmark  & \xmark & \cmark  & \cmark  & \xmark & \cmark  & \cmark  \\
2013\_q2bfsm (T3) & \xmark & \xmark  & \xmark  & \xmark & \cmark  & \cmark  & \xmark & \cmark  & \cmark  \\
lemmings3 (T3)  & \xmark & \xmark  & \xmark  & \xmark & \cmark  & \cmark  & \xmark & \cmark  & \cmark  \\
\bottomrule
\end{tabular}
\label{tab:synthesis_results}
\end{table}

\begin{table}[t]
\footnotesize
\caption{Pass@1, @5, and @10 (Function) for Selected Benchmarks.}
\centering
\setlength\tabcolsep{3pt}
\begin{tabular}{@{} l *{12}c @{}}
\toprule
\multirow{2}{*}{Benchmark} & 
\multicolumn{3}{c}{QWEN Base} & 
\multicolumn{3}{c}{QWEN-HLS} & 
\multicolumn{3}{c}{\textbf{SAGE-HLS}} \\ 
\cmidrule(r){2-4} \cmidrule(r){5-7} \cmidrule(r){8-10}
& @1 & @5 & @10 & @1 & @5 & @10 & @1 & @5 & @10 \\
\cmidrule(r){1-1} \cmidrule(r){2-4} \cmidrule(r){5-7} \cmidrule(r){8-10}
ringer  (T1)          & \xmark & \cmark  & \cmark  & \xmark & \cmark  & \cmark  & \xmark & \cmark  & \cmark  \\
\cmidrule(r){1-1} \cmidrule(r){2-4} \cmidrule(r){5-7} \cmidrule(r){8-10}
countslow  (T2)       & \xmark & \xmark  & \cmark  & \xmark & \cmark  & \cmark  & \xmark & \xmark  & \cmark  \\
truthtable (T2)       & \xmark & \xmark  & \xmark  & \xmark & \cmark  & \cmark  & \xmark & \cmark  & \cmark  \\
\cmidrule(r){1-1} \cmidrule(r){2-4} \cmidrule(r){5-7} \cmidrule(r){8-10}
ece241\_2014\_q5b (T3) & \xmark & \xmark  & \cmark  & \xmark & \cmark  & \cmark  & \xmark & \cmark  & \cmark  \\
ece241\_2013\_q8 (T3) & \xmark & \xmark  & \xmark  & \xmark & \xmark  & \xmark  & \xmark & \cmark  & \cmark  \\
\bottomrule
\end{tabular}
\label{tab:pass_results}
\end{table}

\section{Conclusion}

In this paper, we presented SAGE-HLS, a novel framework that leverages AST-guided fine-tuning of an LLM to generate synthesizable HLS-C code. By converting verified Verilog designs into high-level C/C++ and enriching the training data with their corresponding AST representations (control flow analysis), our approach effectively enhances the alignment between design intent and hardware implementation (and its correctness). Our experimental evaluations on a modified set of VerilogEval benchmark, coupled with constrained simulation, demonstrate that SAGE-HLS improves the synthesizability of HLS-C codes generated by LLM by 41\%, reaching near-perfect synthesis rates, while also boosting functional correctness by more than 35\%, i.e., more than doubling the number of correctly verified cases compared to the baseline model.

\bibliographystyle{IEEEtran}
\bibliography{refs}

\end{document}